\documentclass[useAMS,usenatbib]{mnras}
\usepackage{graphicx,graphics}
\usepackage{amssymb}
\usepackage{amsmath}
\usepackage{epsfig}
\usepackage{float}
\usepackage{array}
\def\ds{[\Delta S]_{1{\rm ms}}}

\newcolumntype{L}{>{\centering\arraybackslash}m{2cm}}
\newcolumntype{K}{>{\centering\arraybackslash}m{1.5cm}}

\title[Modelling   FRB  Population   \&  event   rate  predictions]{On
  modelling  the Fast  Radio  Burst (FRB)  population  and event  rate
  predictions}        \author[Bera       et        al.]        {Apurba
  Bera$^1$\thanks{apurbabera@iitkgp.ac.in},                 Siddhartha
  Bhattacharyya$^1$\thanks{siddhartha@phy.iitkgp.ernet.in},    Somnath
  Bharadwaj$^{1,2}$,   \newauthor   N.  D. Ramesh  Bhat$^{3,4}$   and
  Jayaram   N.   Chengalur$^5$  \\ $^1$Department  of Physics,  Indian
  Institute  of   Technology,  Kharagpur,  India  \\   $^2$Centre  for
  Theoretical  Studies,  Indian  Institute of  Technology,  Kharagpur,
  India\\ $^3$Australian   Research  Council   Centre  of
  Excellence for All-Sky Astrophysics (CAASTRO)\\  $^4$International
  Centre for Radio Astronomy  Research, Curtin University, Bentley, WA
  6102,   Australia\\  $^5$National  Centre for Radio Astrophysics,  Tata Institute
  of  Fundamental  Research,  Pune 411007,  India} \date{Accepted 19 January 2016}
\pagerange{\pageref{firstpage}--\pageref{lastpage}} \pubyear{2016}

\def\LaTeX{L\kern-.36em\raise.3ex\hbox{a}\kern-.15em
  T\kern-.1667em\lower.7ex\hbox{E}\kern-.125emX}

\begin{document}

\label{firstpage}

\maketitle

\begin{abstract}
Assuming that Fast Radio Bursts (FRBs) are 
of extragalactic  origin,  we have developed  a formalism to predict 
the FRB detection rate  and the redshift distribution of the detected
events for a telescope with given parameters. We have adopted  FRB $110220$, 
for which the emitted pulse energy is estimated to be  $E_0 = 5.4\,\times\, 10^{33}\,{\rm J}$,
as  the reference event. The
formalism requires us to assume models for (1)  pulse
broadening due to scattering in the ionized inter-galactic medium - 
we consider two different models for this,  
(2) the frequency spectrum of the emitted pulse - we consider 
a power law model $E_{\nu} \propto \nu^{-\alpha}$ with $-5 \le \alpha \le 5$, 
and  (3) the 
comoving number density of the FRB occurrence rate  $n(E,w_i,z)$ - 
 we ignore the  $z$ dependence and assume a fixed  intrinsic pulse width
$w_i=1 \, {\rm ms}$  for all the FRBs. The distribution  of the emitted pulse energy $E$ 
is modelled through (a) a delta-function where all 
the FRBs have the same energy $E=E_0$, and (b) a Schechter luminosity function where 
the energies have a spread around $E_0$. The models are all normalized using 
the $4$  FRBs  detected    by  \cite{th13}. Our model predictions  
for the  Parkes telescope are all consistent with the inferred redshift
distribution of the fourteen FRBs  
detected there to date. We also find that scattering places an
upper limit on the redshift of the  
FRBs detectable by a given telescope; for the Parkes telescope this is $z\sim 2$. Considering 
the upcoming Ooty Wide Field Array, 
we predict a FRB detection rate of  $\sim 0.01$ to $\sim 10^3$ per day.

\end{abstract}


\section{Introduction}
In   the   recent   past   a    new   class   of   radio   bursts   of
millisecond-duration,  called Fast  Radio  Bursts  (FRBs), have  been
detected     at     the      Parkes     and     Arecibo     telescopes
\citep{lo07,th13,sp14}. The observed pulses show a dispersion index of
$-2.000 \pm  0.006$ and a scattering  index of $-4.0 \pm  0.4$ both of
which are  the signatures  of propagation  through cold  plasma.
All of the FRBs barring FRB $010621$ have been detected at high
  Galactic latitudes ($\mid b\mid > 5^{\circ}$)  and 
 the large  dispersion measure 
(DM $\sim 400-1100\,{\rm pc}\,{\rm cm}^{-3}$)  of  these pulses  
exceed  the  expected Galactic contribution predicted by the NE2001
model \citep{CL02} in  the direction of the bursts  by a  factor of
  $\sim 10-20$  in most  of the cases. This indicates an  
extragalactic origin of the  FRB sources. Note that \citet{loe14} 
have suggested an alternate interpretation where the FRBs may be of Galactic 
origin, however we do not consider this possibility here. 
The observed flux density together with the redshift  inferred from 
the extragalactic  contribution  to the DM  imply  that  an
enormous amount of energy ($\sim 10^{31}-10^{33} {\rm J}$) is released 
 in each burst. Further, the source size of $\sim 100\,{\rm  km}$ 
inferred  from  the  pulse   widths  of  $\sim  1\,{\rm  ms}$   
imply   extreme environments in these  sources. Unfortunately, no counterpart has
yet been  detected in any  other part of the  electromagnetic spectrum,
which makes it  difficult to determine  the physical origin of the 
 FRBs \citep{pe15}.

\begin{table*} FRB 
\centering
\begin{tabular}{l K K K K c K c}
\hline  \hline FRB  & Peak  flux  density (${\rm Jy}$)  & Pulse  width (${\rm ms}$)  & 
Measured Fluence (${\rm Jy\, ms}$) & DM (${\rm pc}\,{\rm cm}^{-3}$) & z & Spectral index  & Reference \\ \hline
FRB $010621^{\star}$ & $0.4$ & $7.8$ & $3.12$ & $746$ & $0.1$ & $-$ & \cite{ke12}\\  
FRB $010724$ &  $30$ &  $4.6$ &  $140$ &  $375$ &  $0.1$ & $-4\pm1$ &\cite{lo07}\\  
FRB $011025$   &  $0.3$  &  $9.4$   &  $2.8$  &  $790$   &  $0.6$ & $-$  & \cite{BB14}\\
FRB $090625$   & $0.5$  &  $4.4$   &  $>2.2$  &  $900$   &  $0.9$ & $-$  & \cite{ch15}\\  
FRB $110220$   &  $1.3$  &  $5.6$   &  $8.0$  &  $944$   &  $0.8$ & $-$ & \cite{th13}\\
\underline{FRB $110523$}   &  $0.6$  &  $<6.3$   &  $3.8$  &  $623$   &  $0.5$ & $-7.8$ & \cite{ma15}\\ 
FRB $110627$  &  $0.4$  &  $<1.4$  &  $0.7$  &  $723$  &  $0.6$  & $-$ & \cite{th13}\\  
FRB $110703$  &  $0.5$  &  $<4.3$  & $1.8$  &  $1104$  &  $1.0$  & $-$ & \cite{th13}\\  
FRB $120127$  &  $0.5$  &  $<1.1$  &  $0.6$  &  $553$  &  $0.5$  & $-$ & \cite{th13}\\  
FRB $121002$  &  $0.4$ &  $2.1,3.7$  &  $1.5$  &  $1629$ &  $1.5$ & $-$ & \cite{ch15}\\
\underline{FRB $121102$} & $0.4$ &  $3.0$ & $1.2$ & $557$ & $0.3$ & $7$ to $11$ &  \cite{sp14}\\  
FRB $130626$ &  $0.5$  &  $3.2$ & $>1.5$  &  $952$ &  $0.9$ & $-$ & \cite{ch15}\\
FRB $130628$ &  $0.9$  &  $1.4$ & $>1.2$  &  $470$ &  $0.4$ & $-$ & \cite{ch15}\\
FRB $130729$ &  $0.1$  &  $23.4$ &  $>3.5$  &  $861$ &  $0.8$ & $-$ & \cite{ch15}\\
FRB $131104$ &  $1.1$  &  $<0.6$ &  $0.6$  &  $779$ &  $0.6$ & $0.3\pm0.9$ & \cite{ra15}\\  
FRB $140514$   &  $0.5$  &  $2.8$   &  $1.3$  &  $563$   &  $0.4$ & $-$  & \cite{pe15}\\ 
\hline
\end{tabular}
\caption{\textbf{Reported  FRBs to  date}: 
This list contains a  total of {\color{blue}sixteen} events among which {\color{blue}\underline{FRB $121102$} and 
\underline{FRB $110523$} were detected at Arecibo and Green Bank Telescope respectively} while the remaining
{\color{blue}fourteen} were all detected at Parkes. The FRB parameters, including the redshifts 
are taken from the published references listed in 
the Table. {\color{blue}\protect\cite{ch15} have not mentioned the  redshift but provide $DM_{MW}$ from which 
we have estimated $z$  using  eq. (\ref{eq:c3}).} 
Note that none of the FRBs have a direct redshift measurement, and the $z$ values in this Table have all been 
inferred from the observed dispersion measure. 
\protect\cite{BM14} have proposed that FRB $010621^{\star}$ probably has a galactic origin in which case 
the estimated redshift is not 
meaningful. However, in our work we assume an extragalactic origin for this FRB and take the estimated 
redshift to be correct.
}
\label{tab:1}
\end{table*}

Several models have  been proposed  for the  source of the FRBs. These 
include  super-massive neutron stars \citep{FR14}, binary neutron star
mergers  \citep{To13},  binary  white  dwarf  mergers  \citep{ka13},
flaring stars \citep{loe14}, pulsar companions \citep{MZ14} and many
more. However,  we are still far  from having enough information 
to validate any of these models. 

Fourteen out  of the  sixteen  FRBs  so far 
(Table~\ref{tab:1}) were detected 
with  the Parkes telescope, nine of  which were discovered by the  
High Time Resolution Universe  (HTRU)  survey.  
The  Parkes  radio  telescope  is  a  fully
steerable single dish telescope of $64 \, {\rm m}$   diameter. The 
Parkes multi-beam receiver has $13$  independent beams each  with
 a narrow field of  view (FoV)  of $14.4'$  (HPBW).  
 The  combination of low system noise ($27\,{\rm K}$) and a
large gain ($G=0.74\,{\rm K/Jy}$) of the primary beam
 makes it one of the  most 
sensitive single dish telescopes currently in operation. 
The HTRU survey  uses L-band  receivers  operating  at 
$\sim 1.4\,{\rm GHz}$ with  a bandwidth of $400\,{\rm MHz}$.

All the FRBs (except FRB $110523$), to date (Table~\ref{tab:1}),  
have been detected in the L-band ($\sim 1-2\,{\rm GHz}$) and their emission  spectrum is  
not  constrained,  though it  appears  that they may  be
consistent with a flat  spectrum. Detections with telescopes operating
at lower frequencies  will   place strong constraints
on   the   spectrum which in turn will yield very important 
constraints on models for the origin of the FRBs.  Currently all that is
available are  constraints on  spectral indices and event rates  of FRBs
 from   non-detections  in   observations  at   lower
frequencies \citep{co14,kar15}. 

The Ooty Radio Telescope (ORT\footnote{http://rac.ncra.tifr.res.in/})
is a parabolic cylindrical reflector of dimensions $530\,{\rm m}\,\times\,30\,{\rm m}$ 
which  operates at a nominal frequency of $\nu_o= 326.5\,{\rm MHz}$ with a 
bandwidth of $4\,{\rm MHz}$.
It has a linear array of  $1056$ half-wavelength dipoles placed
nearly   end-to-end  along   the   focal  line   of  the   cylindrical
reflector.  Since  the  dipoles   are all  oriented  along  the  same
direction,  the telescope is sensitive to  only  a single linear 
polarization   component.  The  ORT currently operates as a
single  antenna which combines  the signal from all the  $1056$ dipoles.  
We refer to this as the  ORT  Legacy system (LS). Work is currently in progress to 
upgrade the ORT  so that it is possible to  combine  different numbers $(N_d)$ 
of successive  dipoles to form many $(N_A)$ smaller individual  antennae
which can functions as a linear interferometric array, the Ooty Wide Field Array
(OWFA; \cite{PS11a,PS11b}; \cite{smc15}).
At completion  we expect to have OWFA Phase I (PI), OWFA Phase II (PII) and the 
LS, all of which can function in parallel, and for which a few 
relevant parameters  are summarized in  Table~\ref{orttab}.  
The large field  of view and reasonably  high sensitivity makes all three versions 
of the ORT (LS,PI and PII) very promising instruments for detecting FRBs. 
The PII, in 
particular, has a FoV  that is $880$ times larger than that of  Parkes while the 
system noise is only five times larger.  While  the two instruments  work 
at different 
frequencies, this comparison gives an idea of the tremendous potential 
of detecting a  large number  of FRBs.  The two other versions (LS and PI) 
will probe smaller FoVs with deeper sensitivity.
We expect the three versions together provide very interesting and useful inputs as to the 
FRB population and the origin of the FRBs.

\begin{table*}
\centering
\begin{tabular}{c c c c}
\hline 
Parameter & ORT Legacy & OWFA Phase I & OWFA Phase II \\ 
\hline
Number  of dipoles  ($N_d$) &  $1056$ &  $24$ &  $4$ \\
Number  of antennae  ($N_A$) &  $1$ &  $40$ &  $264$ \\  
Aperture dimensions
($b\times d$)  & $530\,{\rm m}\times\,30\,{\rm m}$ &  
$11.5\,{\rm m}\times\,30\,{\rm m}$  & $1.9\,{\rm m}\times\,
30\,{\rm m}$  \\ 
Field  of view   &  $0.1^{\circ}\times\,1.75^{\circ}$  &
$4.6^{\circ}\times  1.75^{\circ}$ &  $27.4^{\circ}\times\,1.75^{\circ}$\\ 
Angular Resolution &  $0.1^{\circ}\times\,1.75^{\circ}$  &
$7^{'}\times  1.75^{\circ}$ &  $6.3^{'}\times\,1.75^{\circ}$\\ 
Bandwidth $B$ in ${\rm MHz}$&  $4$ & $18$ & $30$ \\
Spectral Resolution $(\Delta \nu_c)$ in ${\rm kHz}$ & $125$ & $24$ & $48$\\
$\ds$ in ${\rm Jy}$ & $0.343$ & $1.179$ & $2.151$\\ 
\hline
\end{tabular}
\caption{This shows the system parameters for the ORT  Legacy system and the upcoming
  Phase  I and  Phase II  of OWFA. The  aperture
  efficiency ($\eta$) is approximately  $0.6$, the system 
temperature ($T_{sys}$) is $150\,{\rm K}$ for all the three systems and 
$\Delta S(1\,{\rm ms})$ is the $1~\sigma$ noise for incoherent addition of the 
antenna signals  with integration time of $1\,{\rm ms}$. For reference, the Parkes
 telescope has a FoV (HPBW) of $0.23^{\circ} \times 0.23^{\circ}$ and 
$\ds=0.05 \, {\rm Jy}$  in the L-band.
 We note that it is necessary to do 
offline beam forming to achieve the angular resolution quoted here 
 for OWFA Phases I and II. For this paper we have only 
 considered incoherent addition where the angular resolution is the same as 
 the field of view listed here. 
 }
\label{orttab}
\end{table*}

In this paper we assume the FRBs to be of cosmological origin, and 
set up a general framework for predicting the detection  
rate  for a telescope with given parameters.  As mentioned earlier,
very little is known about the FRBs and it is necessary to make 
several assumptions to make progress.  
To this end, we introduce a power law model for  the spectral
energy density  of a FRB and calculate the fluence and pulse width 
that will be observed  accounting for the various propagation effects
including dispersion and scattering in the  inter-galactic medium (IGM). 
We use this to determine  a FRB detection criteria. It is also  
necessary to specify the comoving number density of the FRB occurrence  rate $n(E,w_i,z)$ as a 
function of the pulse energy $E$, its intrinsic width $w_i$, and the redshift 
$z$. We have considered two simple models, both of which assume 
$n(E,w_i,z)$ to be independent of $z$ over the relevant redshift range. 
The models are all normalized to the FRB detection rate 
observed at the Parkes telescope. Finally, we use the entire framework to make 
predictions for the FRB detection rate for the three different versions of the ORT 
(LS, PI and PII) which, in principle,  can  work commensally.

A brief outline of the paper follows. The framework for calculating the 
detection rate is presented in Section 2. Section 3 presents   the  
models for the   FRB population, and  we present  the  detection rates 
predicted for the three versions of ORT in Section 4. The 
summary and conclusions are presented in Section 5.

FRB $110220$ detected  by  \cite{th13} is  the second  
brightest event observed so far (after  the so-called Lorimer  burst;
\cite{lo07}),  and  it is the best  characterized   FRB at present.  
We have adopted FRB $110220$ as  the reference event   for  our entire  analysis.
FRB $110220$ was detected in beam $03$ of the Parkes multi-beam receiver, 
however its exact position relative to the beam center is not known. 
For our analysis we have made the conservative  assumption   that  it  is 
located   close to  the beam  centre, {\it i.e.} the  
intrinsic fluence is  almost  equal  to  the observed fluence. 
For  our calculations we assume that the Parkes has a Gaussian 
beam shape.

It is improtant to note that  currently  no FRB has an independent
redshift measurement, and all the redshifts quoted in Table~\ref{tab:1}
have been inferred from the measured $DM$ which is assumed to be a sum of
three components. 
The  NE$2001$  model \citep{CL02} gives an estimate of the 
Milky Way contribution in the direction of the FRB. The contribution from the 
FRB host galaxy is unknown, and  different authors have used different values for 
this. The residual $DM$, after accounting for these two components, is attributed 
to an uniform, completely ionized  IGM and this is used to infer the FRB's redshift.  
Both the Milky Way ISM  and the IGM are clumpy and turbulent, and the respective $DM$ contributions 
along the actual line of sight to the FRB will differ from the model prediction used to
infer the redshift.   There is further  uncertainity in the inferred redshift 
as there is no  estimate for  the host 
contribution.
It is possible to avoid this last uncertainty to some extent by setting the host
contribution to zero whereby we may interpret the inferred redshift as an
upper limit to the actual redshift of the FRB \citep{KP15}. The various 
uncertainties in the FRB models adopted later in this paper far outweigh 
the uncertainties in the inferred redshifts, and for the present work 
 we have adopted the values quoted in Table~\ref{tab:1}

 We have used $(\Omega_m,\Omega_{\Lambda},\Omega_b,h)=(0.32,0.68,0.04,0.7)$
for the cosmological parameters  \citep{sp03}.

\section{Basic Formalism}
We assume that the spectral energy density $E_{\nu}$ emitted by an  FRB can be
expressed as 
\begin{equation}
	E_{\nu} = E \phi(\nu)
\label{eq:a1}
\end{equation}
where $\phi(\nu)$ is the emission  profile. 
As mentioned earlier, we have used FRB $110220$ as the fiducial event  for our 
analysis. This FRB was inferred to have a redshift $z= 0.8$, for which 
 the Parkes observational frequency  band from $1182\,{\rm MHz}$ to 
$1582\,{\rm MHz}$  corresponds to the frequencies $\nu_a=2128\,{\rm MHz}$ and  
$\nu_b=2848\,{\rm MHz}$ respectively  in the rest frame of the source.

We have used the frequency interval from 
$\nu_a=2128\,{\rm MHz}$  to $\nu_b=2848\,{\rm MHz}$ to normalize the 
emission line profile of all the FRBs such that 
\begin{equation}
	\int_{\nu_a}^{\nu_b}\phi(\nu)d\nu=1 \, .
\label{eq:a2}
\end{equation}
Here $E$ (eq. \ref{eq:a1}) is  the energy emitted by the FRB  in the 
frequency interval $\nu_a$ to $\nu_b$,   and we  henceforth 
refer to  $E$  simply as the ``energy''  emitted by the 
FRB. For reference, the   energy emitted  by FRB $110220$
is estimated to be  $E_0 = 5.4\,\times\, 10^{33}\,{\rm J}$  which 
we use as the fiducial value of 
$E$ throughout this paper. 

We now consider  observations of an FRB  of energy $E$ at redshift  $z$. 
The number of  photons emitted  in the frequency interval  $d\nu_{src}$ 
centred around  $\nu_{src}$ in  the rest  frame of  the source  is
given by
\begin{equation}
  dN_{photon}=\frac{E\phi(\nu_{src})\,d\nu_{src}}{h_p\nu_{src}}
\label{eq:b1}
\end{equation}
where $h_p$ is the Planck constant. The same number of photons 
will be received in the frequency  interval  
$d\nu_{obs}=(1+z)^{-1} \, d\nu_{src}$ centred around the frequency 
$\nu_{obs}=(1+z)^{-1} \, \nu_{src}$  at the observer.
 The fluence  $F_{\nu_{obs}}$ observed in this frequency interval is 
\begin{equation}
  F_{\nu_{obs}}=\frac{dN_{photon}  h_p\nu_{obs}}{4\pi r^2 d\nu_{obs}}.
 \label{eq:b2}
\end{equation}
where $r$  is the comoving distance corresponding to redshift $z$. 
The usual unit of comoving distance is Mpc.
Using eqs. (\ref{eq:b1}) and (\ref{eq:b2}),  we have 
\begin{equation}
  F_{\nu_{obs}}=\frac{E\phi(\nu_{obs}(1+z))}{4\pi r^2}\, .
 \label{eq:b3}
\end{equation}

We now introduce the assumption that the observations are being 
carried out  using a 
telescope with an observational frequency band from $\nu_1$  to $\nu_2$. 
In this  context it is useful to introduce the  average  line profile 
$\overline{\phi}(z)$ defined as   
\begin{equation}
\overline{\phi}(z)=\frac{1}{(1+z)(\nu_2-\nu_1)}\int_{\nu_1(1+z)}^{\nu_2(1+z)}
\phi(\nu) \, d \nu  \,.
\label{eq:b4}
\end{equation}
Further, we also assume that the FRB is located at an angle 
$\vec{\theta}$ relative to the telescope's beam center, and 
use $B(\vec{\theta})$ to denote the normalized beam pattern. 
The fluence that will be observed by this telescope is given by 
\begin{equation}
\overline{F}=\frac{E\overline{\phi}(z)B(\vec{\theta})}{4\pi r^2}.
\label{eq:b5}
\end{equation}

\subsection{Pulse Broadening}
An electromagnetic pulse from cosmological distances gets broadened by
three factors - cosmic expansion, dispersion and scattering, the later
two  due   to  propagation   through  the  ionized   Inter   Stellar  
Medium (ISM) and  the  Inter   Galactic Medium (IGM). The cosmic  
expansion simply broadens the pulse by  a factor of $(1+z)$. The observed 
pulse width  $w$ for an extragalactic event with and intrinsic pulse 
width of $w_i$ is given by
\begin{equation}
w=\sqrt{w_{cos}^2+w_{DM}^2+w_{\rm sc}^2}
\label{eq:c1}
\end{equation}
where  $w_{cos}=(1+z) w_i $, 
$w_{DM}$   and  $w_{sc}$  are  respectively the   contributions to the total 
pulse width  from the cosmologically broadened intrinsic pulse width,the    
residual  dispersion across  a  single  frequency
channel and scattering in the intervening medium.

The frequency dependent refractive index  of the  ionized  components 
of the ISM and IGM causes dispersion of a pulse propagating  through 
it. This dispersion, which has a $\nu^{-2}$ dependence, spreads the 
observed pulse over a large time interval across the entire observational 
frequency bandwidth $B$.  The signal is dedispersed by applying  
appropriate time delays to synchronize the  pulse at all the frequency 
channels across the band. However, it is not possible to correct for 
the dispersion within a single frequency channel width  $\Delta \nu_c$. 
This introduces a residual dispersion  broadening $w_{DM}$ which, under 
the assumption $\Delta \nu_c/\nu_0 \ll 1$,  can be  calculated using 
\begin{equation}
w_{DM}\approx8.3\times 10^6 \, \frac{DM\,  \Delta \nu_c}{\nu_0^3}\,{\rm ms}
\label{eq:c2}
\end{equation}
where  $\nu_0$ is the central frequency of the observation expressed  
in  ${\rm MHz}$ and the dispersion  measure ($DM$)  is  expressed in 
${\rm pc}\,{\rm cm}^{-3}$. Note that the fact that we are holding the 
frequency in the denominator of eq.( \ref{eq:c2}) fixed at the value 
$\nu_0$ instead varying it from channel to channel will introduce  a 
few per cent error in the case of broad band observations. 

As mentioned earlier, the total line  of sight $DM$ has roughly three
contributions  
respectively originating from the Milky Way ($DM_{MW}$), the Inter 
Galactic Medium ($DM_{IGM}$) and the host galaxy ($DM_{Host}$) of the 
source, and  we can write
\begin{equation}
DM=DM_{MW} + DM_{IGM} + DM_{Host} \, .
\label{eq:c3}
\end{equation} 

We can estimate the electron density along different lines of sight
in the  Milky Way by from  the  NE$2001$  model \citep{CL02} and use 
this to calculate $DM_{MW}$ along the line of sight to the FRB.
We use $DM_{MW}=60\,{\rm pc}\,{\rm cm}^{-3}$ as  a representative 
value for  directions away from  the Galactic plane ($b > 5$).
For the host galaxy, we assume that it is similar to the  Milky
  Way with the difference that  we have no idea of the position of the FRB
relative to the  disk of the host  galaxy and we have to allow for the
possibility that the FRB signal reaches us through the disk of the
host galaxy.  We therefore 
expect that on the average  the FRB signal will traverse a larger distance
through the ISM of the host galaxy as compared to the distance it traverses
through the Milky Way, and we use  a slightly larger value 
$DM_{Host}=100/(1+z)\,{\rm pc}\,{\rm cm}^{-3}$ for the host galaxy
contribution. The $(1+z)$  factor here arises due to the cosmological
expansion.  We  estimate   the IGM contribution using  \citep{io03}
\begin{equation}
DM_{IGM}(z)=\frac{3cH_0\Omega_b}{8\pi Gm_P} \int_0^z
\frac{(1+z')dz'}{\sqrt{\Omega_m(1+z')^3+\Omega_{\Lambda}}}
\label{eq:c4}
\end{equation} 
where $m_P$ is the proton mass and the other symbols have 
the usual interpretation. 

\begin{figure}
\centering \input{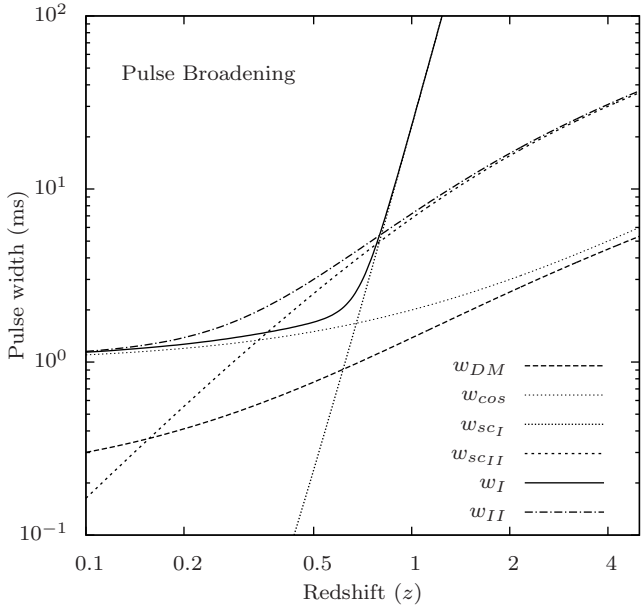}
\caption{The predicted total  pulse  width assuming an FRB of intrinsic 
pulse width $w_i=1 \, {\rm ms}$ located at redshift $z$ observed by the Parkes 
telescope. The subscripts $I$ and  $II$  denote  the   two different scattering  models 
referred to in the text. The different components which contribute to the total 
pulse width are also shown individually.}
\label{fig:1}
\end{figure}

Multipath propagation through the ionized IGM and the  ISM  of 
both the host  galaxy and  the Milky Way  cause  scatter
broadening $w_{sc}$ of the  pulse.  We expect this  to predominantly  
arise from  the IGM  due to a  geometrical effect  known as the 
lever-arm effect \citep{Va76}. The theory of scatter broadening 
in the ionized IGM is not well understood at present, and we  
consider two scattering models to calculate the pulse broadening. 
\begin{enumerate}
\item  \textbf{Scattering Model  I} is  based on  the empirical  fit
\begin{equation}
\begin{split}
\log\:w_{sc}=C_0 &+ 0.15\:\log\:DM_{IGM} \\ &+1.1\:(\log\:DM_{IGM})^2-3.9\:
\log\:\nu_0
\label{eq:c5}
\end{split}
\end{equation} 
with $C_0=-6.46$   given by \cite{bh04} for  pulsars in  the ISM  of  
our Galaxy. We have used $C_0=3.2$ to rescale this  for  scattering  
in  the  IGM. This value of $C_0$ is based on the assumption that the 
reference event FRB $110220$ has an intrinsic pulse width of  
$w_i=1 \, {\rm ms}$.
Eq. (\ref{eq:c2}) predicts $w_{DM}=0.17  \, {\rm ms}$ for $z=0.8$, 
and we have set the value of $C_0$ so that eq. (\ref{eq:c5}) gives 
$w_{sc}=5.3  \, {\rm ms}$ required to match the observed pulse width 
$w=5.6 \, {\rm ms}$  (Table~\ref{tab:1}). Note that in eq.( \ref{eq:c5}), 
we use  $\nu_0$ in ${\rm MHz}$, $w_{sc}$ in ${\rm ms}$,  $DM$  in 
${\rm pc}\,{\rm cm}^{-3}$ and $\log$  denotes  $\log_{10}$.

\item \textbf{Scattering Model II} is based on the temporal 
smearing equation for IGM turbulence  given by  \cite{MK13} 
\begin{equation}
w_{sc}(z)=  \frac{k_{sc}}{\nu ^4  Z_L} \int_0^z  D_H (z')dz'  \int_0^z
(1+z')^3 D_H (z') dz'
\label{eq:c6}
\end{equation}
where 
\begin{align}
D_H  (z) =& (\Omega_{m}(1+z)^3+\Omega_{\Lambda})^{-1/2}, \\
Z_L=& (1+z)^2    \left[(1+z)-\sqrt{z(1+z)}\right]^{-1}
\end{align}
and we use 
$k_{sc}=8.5\,\times\,10^{13}\,{\rm ms \,  {MHz}^4}$ for the 
normalization constant. As with Scattering Model I, the value of 
the normalization constant is set to reproduce the observed pulse 
width of FRB $110220$ assuming that it has an intrinsic pulse width  
$w_i=1 \, {\rm ms}$. 

For both the scattering models that  we have considered here, 
the value  of the normalization constant  would change  if we assume 
a different intrinsic pulse width for FRB $110220$. We have tried out  
$w_i=0.5$ and $2 \, {\rm ms}$ in order to assess how this would affect 
the results of our analysis. 

Note that Scattering Model I  is based on an
empirical  fit which  is observationally well established 
within the ISM of our  Galaxy. 
Given the high-DM part of the model is largely constrained by measurements of 
pulsars in the Galactic plane, it is effectively a representation of turbulence and clumpiness 
in the ISM. The nature of turbulence may be different for IGM and hence it is not clear  
whether the same fit can be rescaled to correctly quantify  IGM scattering. In contrast, 
Scattering Model  II is  based entirely on  a theoretical model 
for the  IGM scattering.  This model, however, has not been 
observationally  verified. Given our present lack of knowledge, 
we have used the two different scattering models to estimate the possible 
impact on the pulse width.  
\end{enumerate}

Figure (\ref{fig:1}) shows the total pulse width $w$ (eq. \ref{eq:c1})
corresponding to an FRB of intrinsic pulse width $w_i=1 \, {\rm ms}$
located at a redshift  $z$ observed by the Parkes telescope 
for which $\nu_0 = 1382\,{\rm MHz}$ and $\Delta \nu_c = 390\,{\rm kHz}$. 
Recollect that both the scattering models are normalized using 
FRB $110220$ assuming $w_i=1 \, {\rm ms}$ for this event, and  therefore 
both the scattering models predict  $w=5.6\,{\rm ms}$
at $z=0.8$. We see that the  residual dispersion measure $w_{DM}$ makes 
a very insignificant contribution to $w$
at all redshifts. Scatter broadening  is not very important at low redshifts where 
we have $w \approx w_{cos}$. The total pulse width is dominated by scatter broadening 
 at large redshifts. 
For  Scattering Model I the total pulse width is dominated by 
$w_{sc_I}$ at $z  \approx 0.5$, and 
 $w_{sc_I}$  increases sharply for $z \ge 0.5$.
For  Scattering Model II  $w_{sc_{II}}$ starts making a significant contribution 
to $w$ at $z  \approx 0.2$, however it dominates the total 
$w$ only  for $z \ge 0.4$. Unlike  Scattering Model I,   $w_{sc_{II}}$
increases relatively gradually with $z$. For both the scattering models, 
the total pulse width is considerably in excess of $w_{cos}$ for FRBs 
with  $z \ge 0.5$. We have repeated the entire exercise for $w_i=0.5$ and
 $2 \, {\rm ms}$, {i.e.} the intrinsic pulse width of the observed FRB 
and the normalization of the scattering model are both changed. 
The results, which are very similar to those shown in Figure (\ref{fig:1}) for 
$w_i=1 \,  {\rm ms}$, are not shown here. We find that the qualitative features
of the total pulse width as a function of $z$ 
are not very different if we change the value of $w_i$. Scatter broadening 
starts to dominate at $z  \approx 0.5$, and the total pulse width is 
considerably larger than $w_{cos}$ for $z \ge 0.5$.

\subsection{Detection criteria and detection rate}
Considering  an FRB of energy $E$ and intrinsic pulse width $w_i$ 
located at redshift $z$,  we have  till now  discussed how to calculate the fluence 
$\overline{F}$ (eq. \ref{eq:b5}) and the pulse width $w$ (eq. \ref{eq:c1}) that will be observed
by a given telescope. We now discuss  the criteria for this particular  FRB to be detected by 
the given telescope.

The detection criteria is decided by the  telescope's system noise. The
r.m.s. flux density fluctuation  $\Delta S$ is given by 
\begin{equation}
\Delta S=\frac{T_{sys}}{G\sqrt{\Delta t \, B \,  N_{pol}}} \equiv 
\sqrt{\frac{1\;{\textrm ms}}{w}} \times \ds 
\label{eq:d1}
\end{equation}
where  $G$ is the antenna gain of the primary beam,  $T_{sys}$ is the telescope's system  temperature, 
$N_{pol}$ is  the  number of  polarizations  the telescope  detects  and $\Delta t$ 
is the  integration time. We  assume an  integration time
equal to the observed pulse width $w$. Since the observed FRB pulse widths are of
the order of a few milliseconds, 
it is then convenient   to express $\Delta S$ (eq. \ref{eq:d1}) in terms of  
$w$ and  $\ds$ which  is  the r.m.s.  noise for   $\Delta t=1\,{\rm ms}$. 

\begin{figure*}
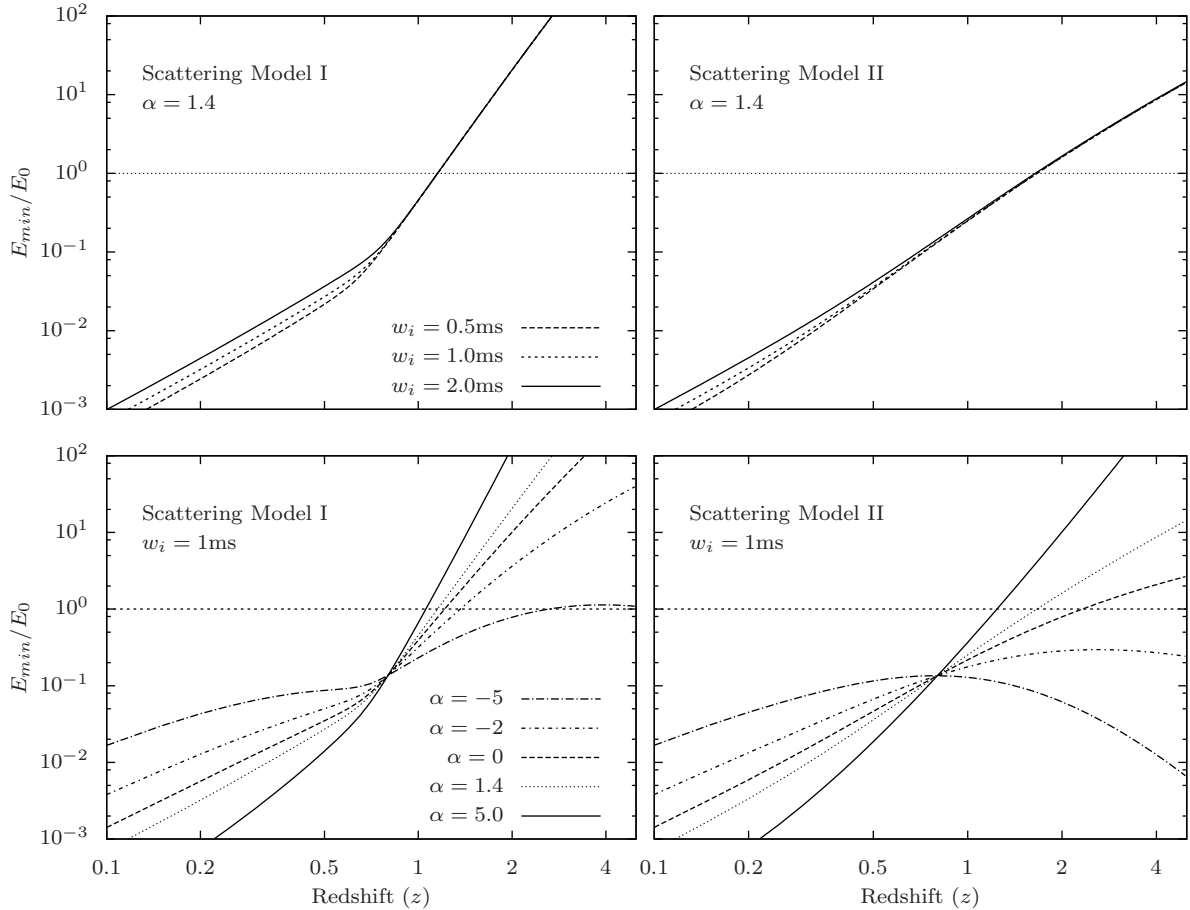

\centering
	\input{plots/parkes/bhatscatter_eminalp.tex}
    \input{plots/parkes/lensscatter_eminalp.tex}
	\input{plots/parkes/bhatscatter_eminw.tex}
    \input{plots/parkes/lensscatter_eminw.tex}
\caption{ The minimum  energy $E_{min}$ for  an FRB  at redshift $z$ 
to be detected at the Parkes  telescope assuming that the source
  is located  at the beam centre. The left and right panels are for 
Scattering Models I and II respectively. The upper panels consider 
different values  of the intrinsic pulse widths $w_i$ with 
spectral index  $\alpha=1.4$ fixed, while the lower panels consider  
different values of $\alpha$ with $w_i=1 \, {\rm ms}$ fixed.}
\label{parcrit}
\end{figure*}

An FRB   with  average  observed  flux density $\overline{S}=\overline{F}/w$ will result in 
a  detection if 
\begin{equation}
\frac{\overline{S}}{\Delta S} =\frac{\overline{F}}{w \, \Delta S} \ge n \,,
\label{eq:d2}
\end{equation}
where $n$ is  the minimum  signal to  noise ratio required for a detection. 
The same criteria can be expressed in terms of a limiting fluence 
 $F_l=n \, \times (1\,{\rm ms})\times \ds$ as 
\begin{equation}
\overline{F} \times \sqrt{\frac{1\;{\textrm ms}}{w}}
\geq F_l
\label{eq:d3}
\end{equation}
\citet{th13} have only considered events with a signal to noise ratio greater than 
nine as a detection, following these authors  we have used $n=9$ for the Parkes telescope.

The detection criteria (eq. \ref{eq:d2}) combined with  (eq. \ref{eq:b5}) implies 
a minimum energy 
\begin{equation}
\label{lmin}
E_{min}=\frac{4\pi r^2 F_l }{\overline{\phi}(z)B(\vec{\theta})} \, \sqrt{\frac{w}{1\;{\textrm ms}}}
\end{equation}
for a telescope to detect  an FRB  at a redshift  $z$  and a sky position
$(\vec{\theta})$ relative to  the telescope's beam centre. A telescope's primary beam 
pattern $B(\vec{\theta})$, the antenna  gain $G$ of the primary beam
and system temperature $T_{sys}$
will, in general, vary as the telescope is pointed to different parts of the 
sky. To simplify the analysis, we  have assumed    these 
telescope parameters to be constant. 
 
The  number of  FRB events at redshift  $z$  per unit  time (in  the source
frame) per  unit comoving volume  with energy  in the
range  $E$ to  $E+dE$ and  intrinsic  pulse width  in the range  $w_i$  and
$w_i+dw_i$ can be expressed  as
\begin{equation}
dN=n(E,w_i,z)\, dE \, dw_i.
\end{equation}
where $n(E,w_i,z)$ is the  
comoving number density of the FRB occurrence  rate.  

For an observation  time $T$ with a given telescope, the number of events detected  
($N_{det}$) is expected to be
\begin{equation}
\begin{split}
N_{det}(T)=&T \int dz \frac{dr}{dz} \left(\frac{r^2}{1+z}\right) \\ 
           &\int d\Omega\int dw_i \int_{E_{min}(z)}^{\infty} dE\:\:n(E,w_i,z).
\label{eq:d5}
\end{split}
\end{equation}
We use eq. (\ref{eq:d5}) to predict  the FRB detection rate for any given 
telescope. Here it is assumed that the telescope has a sampling
  time  $\le 1\,{\rm ms}$ so as to be able to  resolve the FRB. 
The factor of  ($1+z$) in the denominator arises from  the fact that a
time  interval of  $T$ in  the observer's  frame corresponds  to a  time
interval of  $T/(1+z)$ in  the source frame.  The quantity  within the
square brackets gives the redshift distribution of the detected events.
We may interpret the latter as the detection rate with the FRB source originating 
in the redshift interval $z$ to $z+dz$. 

\section{Modelling the FRB Population}
The basic formalism introduced in the previous section requires 
the FRB emission line profile $\phi(\nu)$ and the  comoving number density of  the 
FRB occurrence  rate $ n(E,w_i,z)$ as inputs in order to predict the FRB detection 
rate for any given telescope.  With only sixteen FRB events detected to date, we do not as yet 
have any  established models and  for our  work we  assume very simple models for  
for  these two quantities. We discuss these models below. 

The spectrum of  the FRB emission is very poorly  constrained at present, 
the  detection so far being all (except FRB $110523$) in the L-band. 
 The  high
observed flux  suggests that  the emission  mechanism is  coherent for
which we expect a negative spectral index. Here we have assumed a simple
power  law spectrum  $S_{\nu} \propto  \nu^{-\alpha}$
with  a  spectral index $-\alpha$ vary from $-5$ to $+5$ for which we have 
the normalized  (eq. \ref{eq:a2}) emission profile 
\begin{equation}
\phi(\nu)=\left[\frac{1-\alpha}{\nu_b^{1-\alpha}-\nu_a^{1-\alpha}}\right]
\nu^{-\alpha}
\label{eq:f1}
\end{equation}
and the average line profile  has the form
\begin{equation}
  \overline{\phi}(z)=K(\alpha)(1+z)^{-\alpha}
\end{equation}
where $K(\alpha)$ is
\begin{equation}
K(\alpha)=\frac{1}{\nu_2 - \nu_1}\left[\frac{\nu_2^{1-\alpha}-\nu_1^{1-\alpha}}
{\nu_b^{1-\alpha}-\nu_a^{1-\alpha}}\right] \,.
\end{equation}
Although non-detection in 
searches at different  wavelengths give constraints on the spectral index, we
consider the range $-5 \le \alpha \le 5$ for completeness. 

The  minimum energy $E_{min}$ (eq. \ref{lmin}) can now be expressed as 
\begin{equation}
E_{min}=\frac{4 \pi r^2 F_l }{B(\vec{\theta})}
\left[ \frac{(1+z)^{\alpha}}{K(\alpha)}\right] \,  \sqrt{\frac{w}{1\;{\textrm ms}}} \,.
\end{equation} 

\begin{figure*}
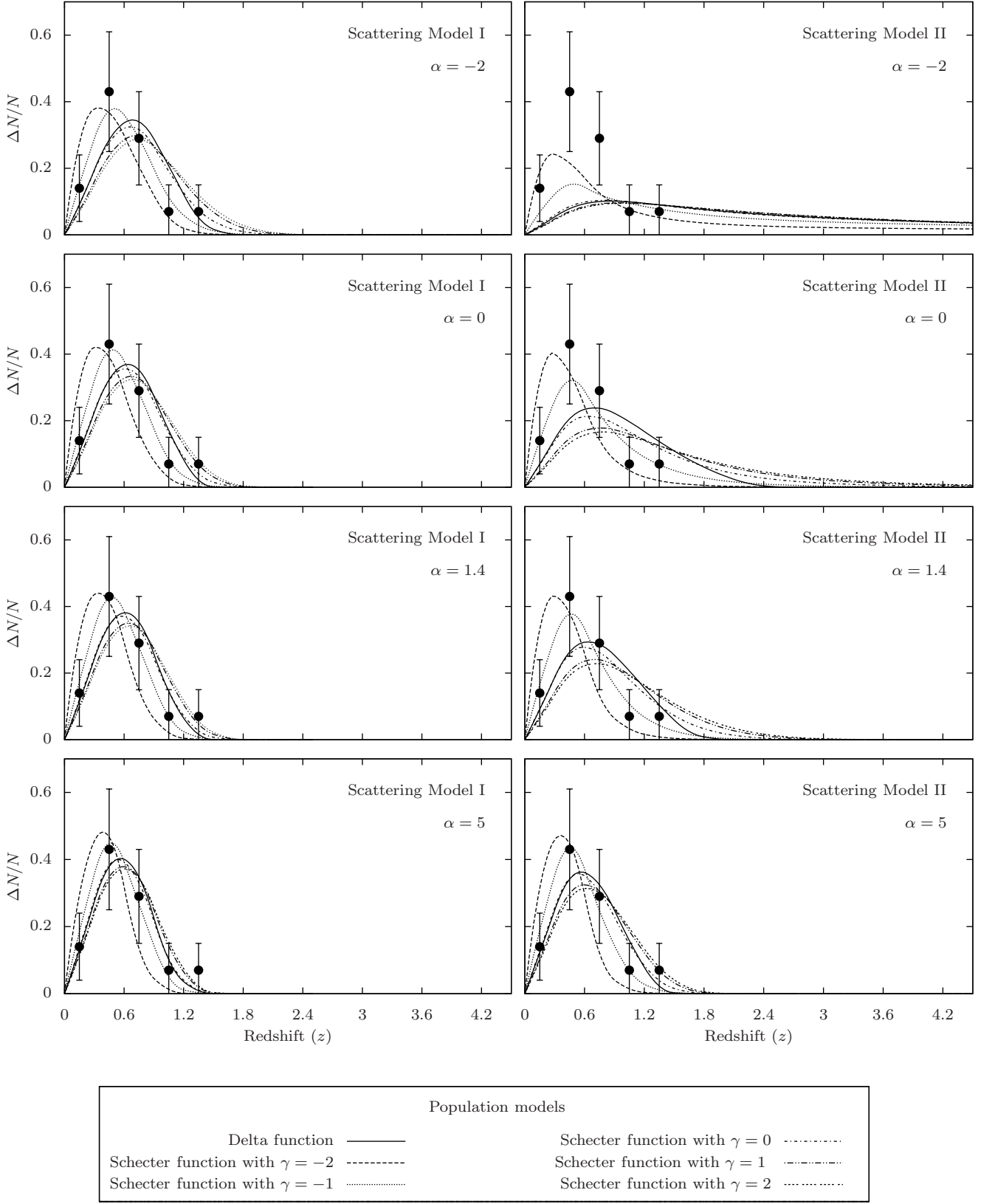

\centering
	
	\input{plots/parkes/bhatdist2m.tex}
	\input{plots/parkes/lensdist2m.tex}
	\input{plots/parkes/bhatdist0.tex}
	\input{plots/parkes/lensdist0.tex}
	\input{plots/parkes/bhatdist1p4.tex}
	\input{plots/parkes/lensdist1p4.tex}
	\input{plots/parkes/bhatdist5.tex}
	\input{plots/parkes/lensdist5.tex}
	
	\input{plots/parkes/notation.tex}
\caption{The data points show the redshift  distribution $\Delta N/N$ of the 
  fourteen  FRBs detected at the  Parkes telescope, the data has been binned
with $\Delta z=0.3$, $\Delta N$ is the number of FRBs in each bin 
and $N$ is the total number of FRBs. The error bars show the $1-\sigma$ 
Poisson errors for the data. The theoretical predictions 
for the different models are shown as continuous curves.  These  curves show 
$N^{-1} \, (dN/dz)$ normalized so that the total area is same for all.
Unfortunately, the number of FRBs which have been detected to date is 
too small to place meaningful constraints on the models which we have considered.}
\label{pardistfig}
\end{figure*}

Figure \ref{parcrit} shows  $E_{min}$ as a function of $z$ assuming 
the FRB to be located at the center of one of the beams of  the Parkes telescope 
($B(\theta)=1$). We expect the FRBs to typically have an energy $E \sim E_0$, and we 
have shown $E_{min}$ in units of the reference energy $E_0$. The upper panels 
show  the results for three different values of the intrinsic pulse width $w_i$
with $\alpha=1.4$ fixed, while the bottom panels shows the results for 
three different values of $\alpha$ with $w_i=1 \, {\rm ms}$ fixed.  We do not find 
a very big  difference  if the value of $w_i$ is changed, however the results 
vary considerably if the value of $\alpha$ is changed. 
We first discuss only the positive values of $\alpha \, ( \ge 0)$.
For Scattering Model I, we find that the value 
of  $E_{min}$ increases sharply for $z \ge 1$  due to the steep
increase in the pulse  width  (Figure \ref{fig:1}). 
The value of $E_{min}$ increases more gradually in   Scattering Model II. 
In all cases the main  feature is that $E_{min}$ increases with redshift and exceeds $E_0$ in the 
range $1 \le z \le 2$. Assuming that the FRBs have energy $E \sim E_0$, 
this imposes a cut-off redshift $z_c$ such that  observation with the Parkes
telescope  are only sensitive to FRBs with $z \le z_c$. 
We see that  the value of $z_c$ is largely insensitive to $w_I$, however  it shifts to smaller
$z$ if $\alpha$ is increased from $0$ to $5$. In all cases 
we find $ 1 \le z_c \le 2$ for the Parkes telescope. For $\alpha \ge 0$, 
our models predict that we do not expect the Parkes telescope to detect FRBs 
with $z> 2$, 
consistent with the observations summarized in Table~\ref{tab:1}. 
Next, considering the negative values of $\alpha$ we find that the results are quite 
different from those for  $\alpha \ge 0$.  The difference is particularly pronounced for 
Scattering Model II where we see that  the value of  $E_{min}$ decreases with $z$ for $\alpha=-5$.
In this case  we do not have a cut-off redshift and we expect the Parkes telescope to 
detect FRBs out to arbitrarily high redshifts, a prediction that is  inconsistent  with 
the observations summarized in Table~\ref{tab:1}. A similar problem also arises for 
$\alpha=-2$, however it is not as severe as for $\alpha=-5$.  Based on these findings, 
we have restricted the subsequent analysis to $\alpha$ values in the range $-2 \le \alpha \le 5$.

We now shift our attention to  the  comoving number density of  the 
FRB occurrence  rate $ n(E,w_i,z)$. Since the cut-off redshift $z_c$ 
for FRB detection,  at least for the  Parkes telescope,  does not  depend    
on   the  intrinsic  pulse  width $w_i$ (Figure \ref{parcrit}) we assume that 
all the FRBs  have  the same  intrinsic pulse  width of $w_i=1 \, {\rm ms}$. 
Further, since  we expect all  the detected FRBs to be within $z \le 2$, as assume
that $ n(E,w_i,z)$ is constant over the limited redshift range of our interest. 
The function $ n(E,w_i,z)$ is now just a function of $E$, and we have considered two
simple models for the $E$ dependence. 

\begin{enumerate}
\item  \textbf{The delta-function model}  where all  the FRBs emit the same energy 
$E_0$ and 
\begin{equation}
n(E,w_i,z)=n_0 \,  \delta (E-E_0) \,.
\label{eq:e1}
\end{equation}
\item \textbf{The Schechter luminosity function  model} where  the FRB energies
have a spread, the energy distribution being given  by 
the Schechter function 
\begin{equation}
n(E,w_i,z)=\frac{n_0}{E_0}\left(     \frac{E}{E_0}\right)    ^{\gamma}
\exp\left( -\frac{E}{E_0}\right) \,.
\label{eq:e2}
\end{equation}
We consider both positive and negative values of the exponent $\gamma$. The negative values of
 $\gamma$ require a lower cut-off  to  make the  distribution  normalizable.  We have 
considered  a cut-off energy of $E_0/100$ for our analysis. 
\end{enumerate}

We have used the FRBs observed by the Parkes telescope to determine 
the value of the normalization constant $n_0$ which is a free parameter 
in  both the models for $n(E,w_i,z)$.  Though there are fourteen FRBs 
detected  at the Parkes telescope, it is not possible to use all of them 
to calculate an event rate because the exact duration of the observation
is not  known. The  four  FRBs  detected    by  \cite{th13}  correspond to  an  effective
observation time of $298$ days with  a single beam of the Parkes telescope,
and we have used the inferred detection rate along with eq. (\ref{eq:d5}) to determine the value of
 $n_0$. Note that \citet{ch15} have estimated a slightly lower FRB occurance rate
than \citet{th13} but they are consistent with each other within 1-$\sigma$ uncertainities.

 As noted earlier, the  FRB distribution predicted for Parkes extends to arbitrarily large redshifts for negative values of $\alpha$. We see this in the topmost  right panel of Figure \ref{pardistfig} 
 ($\alpha=-2$ and Scattering Model II) where the FRB predictions do not fall of even at $z > 4$. 
This  is even more severe for $\alpha=-5$ which has not been shown  here.   This poses a problem for 
the $z$ integral in eq. (\ref{eq:d5}), and it is necessary to assume an  upper limit  to obtain a finite prediction.  We have assumed an upper limit of $z=5$ for calculating $n_0$. 
 While the upper limit $z=5$ has been introduced here for mathematical convenience, 
we may interpret this as the redshift beyond which the FRB population ends abruptly. 
Finally, we note  that the non-detection  of any FRBs  in the search by \cite{rn15} 
seem to indicate that the  FRB rate  is possibly a  factor of  $3$  to $5$ times smaller 
than that inferred from these four Parkes FRBs.

Redshift estimates are available  for all 
the fourteen FRBs detected  by the Parkes telescope  (Table~\ref{tab:1}). 
The redshift distribution of the detected FRBs provides an independent 
constraint on any model for the FRB population. We now compute the 
redshift distribution predicted by our models ( eq. \ref{eq:d5})
and compare these with the observed redshift distribution of the 
fourteen Parkes FRBs. Six of the fourteen 
Parkes FRBs are in the redshift range $0.4 \le z \le 0.6 $. We see that most of the models
also predict redshift distributions  which peak around  the same  $z$ range.
Considering first the delta-function model where all the FRBs have the same energy $E_0$, 
we see that the redshift distribution extends out to larger redshifts in  Scattering Model II
as compared to  Scattering Model I.  The redshift distribution also extends out to larger
redshifts if the  value of $\alpha$ is decreased. Both of these effects can be understood in
terms  of the cut-off redshift $z_c$ introduced while discussing Figure \ref{parcrit}. 
The  Schechter luminosity function  introduces a spread in the FRB energies.  
The  relative abundance of low energy  FRBs increases if the value of  $\gamma$
is reduced, and we see that the entire redshift distribution 
shifts to lower redshifts for negative values of  $\gamma$.

We see that most of the models considered here are
roughly consistent with the redshift distribution of the
observed FRBs (Figure \ref{pardistfig}). As mentioned
  earlier, the predicted FRB distribution extends beyond $z=4$ for
  $\alpha=-2$ and Scattering Model II whereas the observed redshift
  distribution falls off well within $z=1.5$. Unfortunately, the
number of FRBs which have been detected to date is too
small to place meaningful constraints on the models which we have
considered.  However, we anticipate that larger numbers of FRBs will
be detected in future and it will be possible to constrain both the
scattering models as well as the models for the FRB population using
the redshift distribution of the detected events. 

\section{Predictions for OWFA}
We now use the formalism and the various models presented in the earlier 
parts of this paper to study the prospects of detecting FRBs with the ORT 
Legacy System (LS) and  the two different phases of OFWA  (PI and PII). 
In Phase I, the signal from  $N_d=24$ successive  dipoles are combined to  form an
individual antenna, and there are a  total of $N_A=40$ such antennas. 
We have $N_d=4$ and $N_A=264$ for Phase II. The aperture dimensions 
and other details are tabulated in Table~\ref{orttab}.

Each phase of OWFA has  $N_A$ antennas which can be operated together as a linear
radio-interferometric  array. However, for the present analysis we consider
a simpler situation where the signals from the  $N_A$ antennas are incoherently
added. A more detailed analysis using the full beam forming capability of OWFA
will be presented in a later paper. 
Consequently, the field of view is the same as that of a single antenna 
(given in Table~\ref{orttab})  but the r.m.s. flux density fluctuation  is
reduced to  $\ds/\sqrt{N_A}$. The ORT LS and OWFA PI and PII all have
anisotropic  
beam patterns which we have parametrized as 
\begin{equation}
B(\vec{\theta})={\rm sinc}^2\left(  \frac{\pi  d \theta_x}{\lambda}  \right)
{\rm sinc}^2\left( \frac{\pi b \theta_y}{\lambda} \right)
\label{eq:ow1}
\end{equation}
\citep{AB14}
where we have the antenna aperture $b \times d$ (Table~\ref{orttab})
and $\lambda$ is the observing wavelength. Here we have used the 
flat-sky approximation, and $(\theta_x,\theta_y)$ are the components 
of the vector $\vec{\theta}$ on the plane of the sky. We note that 
the flat sky approximation does not hold for Phase II which has a 
very large field of view, however this is justified by the fact 
that the error introduced by this assumption 
is  small compared to the other uncertainties in our modelling of 
the scattering and the FRB population.

\begin{figure*}
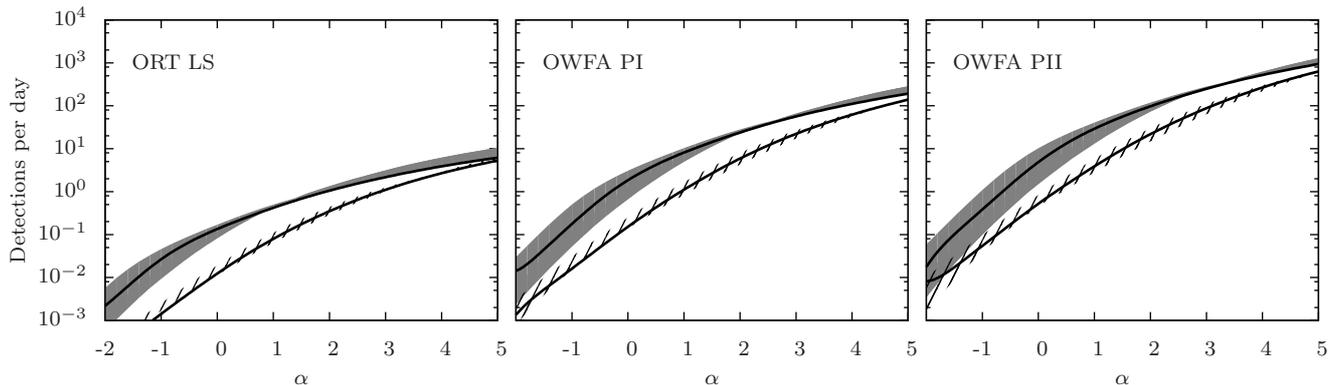

\centering 
\input{plots/ortdet/ort1024totaldet.tex}
\input{plots/ortdet/ort1024ph1indet.tex}
\input{plots/ortdet/ort1024ph2indet.tex}
\caption{The expected FRB detection rates as a function of $\alpha$
 for  ORT LS, and  OWFA PI and PII assuming $1024$ frequency
channels spanning the bandwidth $B$ given in Table~\ref{orttab}. 
The  gray  shaded and hatched regions correspond to 
Scattering Model I and  II  respectively, the solid curves
passing through these regions show predictions for the 
delta-function FRB population model while the boundaries of 
the regions  enclose the curves corresponding to all the other 
models considered in Figure~\ref{pardistfig}.}
\label{detin}
\end{figure*}

Our predictions for the FRB detection rate are shown in 
Figure~\ref{detin}. We have taken minimum signal to noise ratio $n=10$ for these predictions. 
We see that the predicted  detection rate 
is highest for PII which has the largest field of view and the largest 
frequency bandwidth. 
We expect to detect somewhere between $\sim 0.01$ to  $1,000$ FRBs 
per day with PII, depending on the  spectral index of the FRB emission.
We have a higher detection rate  for larger positive values of $\alpha$, 
the detection rates are also higher for Scattering Model I 
as compared to Model II. The predicted detection  rates fall by 
roughly an order of magnitude for PI, and 
roughly two orders of magnitude for LS as compared to PII. 
Considering $\alpha=1.4$  which has been proposed to be the most likely
value for the coherent FRB emission  \citep{lo13},  we  expect to detect 
$\sim 10$  to   $\sim 100$ FRBs   per day with PII. This  is quite
 encouraging even if  we
take into account the fact that \citet{rn15} have suggested that the FRB occurrence 
rates  estimated from the FRBs detected by \citet{th13} may be a  factor of $3$ to $5$
larger than the actual FRB occurrence rate. 
 
\section{Summary \& conclusion}
The source of the FRBs is still largely unknown. Assuming the FRBs to be 
of extragalactic origin,  we have developed  a formalism to predict 
the FRB detection rate (eq. \ref{eq:d5}) 
 and the redshift distribution of the detected
events for a telescope with given parameters. 
We have adopted  FRB $110220$ (Table~\ref{tab:1}) as  the reference event  
for  our entire  analysis. None  of the FRBs detected  to date
have a  direct redshift measurement, and the redshifts of all the detected
FRBs  (Table \ref{tab:1})  have  been  inferred from the observed $DM$s. 
The value of the inferred redshift depends on the Milky Way and host galaxy
$DM$ contribution assumed in the analysis, and different authors have
assumed different values.  For our analysis we have
adopted the inferred $z$ values from the references listed in Table
\ref{tab:1}.  In contrast, the redshift $z$ in our analytic calculations and
in Figures \ref{fig:1}, \ref{parcrit} and \ref{pardistfig} refer to the
actual cosmological redshift of the FRB which is unaffected by our assumptions
for $DM_{MW}$ and $DM_{Host}$.  The assumed values only affect the observed
pulse width $w$ through eqs. (\ref{eq:c1}), (\ref{eq:c2}) and (\ref{eq:c3}). 
Further, the $DM$ makes a subdominant contribution to the total pulse width $w$
for the entire range considered here (Figure \ref{fig:1}), and consequently our
predictions are largely unaffected by  $DM_{MW}$ and $DM_{Host}$.
 
The  FRB pulse width  plays an important role in determining the 
detection rates.  At present we lack adequate
understanding  of pulse broadening due to scattering in the ionized IGM,
and we consider two different alternatives to model this. Scattering Model I
is based on an observational fit  given by \cite{bh04} for  pulsars in the ISM
of   our Galaxy, we have extrapolated this for  FRB pulse broadening in the IGM.
In contrast, Scattering Model II is  based  on a theoretical calculation 
given by  \cite{MK13}, and it has no observational confirmation at present.  
Both the scattering models are normalized to reproduce the observed pulse 
width of FRB $110220$,  assuming that it has  an intrinsic pulse width of  
$w_i=1 \, {\rm ms}$. For the Parkes telescope, 
we find that in both the models scatter broadening 
starts to dominate the total pulse width  at $z  \approx 0.5$
(Figure \ref{fig:1}). In Model I, 
the total  pulse width increases steeply for   $z > 0.5$ whereas a more
gradual increases is predicted by Model II. We also find that the $z > 0.5$
behaviour is not  significantly modified if we assume  an intrinsic pulse width of
$w_i=0.5 $ or $2 \, {\rm ms}$ for FRB $110220$. The cosmological broadening of 
the intrinsic pulse width dominates at lower $z$.

The total energy in the FRB pulse and its spectrum also play an important role 
in determining the FRB detection rate. We have introduced the FRB energy $E$ 
and the FRB  emission profile $\phi(\nu)$ (eq. \ref{eq:a1}) to model 
the FRB  energy spectrum. For our work we have assumed a power law 
$\phi(\nu) \propto \nu^{-\alpha}$ (eq. \ref{eq:f1}) where
  $\alpha$ is the  (negative) spectral index. 
In this paper we have presented results for $\alpha$ values in the range 
$-5 \le \alpha \le 5$, however most of the analysis is restricted
to $\alpha \ge -2$. 

It is necessary to model the FRB population in order to 
make predictions for the detection rate. 
We have quantified the FRB population through $n(E,w_i,z)$ which gives  the 
comoving number density of  the FRB occurrence rate. For our work we have assumed
that  $n(E,w_i,z)$ does not vary  with $z$ over the limited redshift range of our 
interest. Further, all the FRBs are assumed to have the same intrinsic pulse width
$w_i=1 \, {\rm ms}$.  For the $E$ dependence we have adopted the simplest delta-function 
model where all the FRBs have the same energy $E_0 = 5.4\,\times\, 10^{33}\,{\rm J}$
which is  the  estimated  energy for FRB $110220$. We have also considered 
a set of models where the $E$ values have a spread around $E_0$. In this case
the $E$ distribution has the form of a  Schecter luminosity function (eq. \ref{eq:e2}).
We present results for both negative and positive values of $\gamma$. The models are all 
normalized to reproduce the event rate corresponding to the 
four  FRBs  detected  at Parkes   by  \cite{th13}.   

We have calculated the FRB redshift distribution predicted by our models 
for observations with  the Parkes telescope. The predictions indicate that 
we do not expect to detect FRBs with redshifts $z > 2$ with the Parkes 
telescope (Figure \ref{parcrit}).  The redshift distribution 
(Figure \ref{pardistfig}) peaks in the  range $0.4 \le z \le 0.6$  for most of the models 
which we have considered. We find that most of the models that
 we have considered are  
consistent with the redshift distribution of the fourteen FRBs observed by
the Parkes telescope. However, some of the models
with $\alpha \le -2$ predict a redshift distribution that extends beyond 
$z\ge 4$ while the observed FRB distribution is restricted within $z \le 1.5$. 
The number of FRBs observed to date is to small 
to conclusively constrain the models which we have considered here. Our
prediction  however indicate that it will be possible to distinguish between
the different models when more FRB data becomes available in future. 

Finally, we have used the formalism and the different models presented here 
to predict the FRB detection rate expected at the ORT LS and the OWFA PI and PII. 
The main point to note here  is that OWFA PII has a field of view which is
 $880$ times larger than the  individual beam of the  Parkes telescope where 
most of the  FRBs have been  detected. Further, the existing FRBs have all been
 detected in the  L-band ($\sim 1-2\,{\rm GHz}$) whereas ORT and OWFA operate around 
$326.5 \, {\rm MHz}$. 
We predict that we expect to detect somewhere between $\sim 0.01$ to  $1,000$ FRBs 
per day with PII, depending on the  value of $\alpha$.  The upcoming OWFA PII
holds the potential of dramatically increasing the population of  detected FRBs,
thereby opening a new window to unravel the source  of these mysterious  events. 
We plan to present a detailed treatment of the predictions for ORT and OWFA in a 
subsequent work.

\label{lastpage}

\end{document}